# Progress in Decompositional Electromagnetic Analysis of Digital Interconnects

Yuriy Shlepnev

*Simberian Inc.*
*615 Hampton Dr., Unit B306, Venice, CA 90291, USA*
shlepnev@simberian.com

*Abstract* — **In the realm of PCB and packaging interconnect design, electromagnetic analysis tools have transitioned from optional to essential over the last two decades, as data rates soared beyond 6 Gbps. Today, with standard data rates eclipsing 6 Gbps and reaching thresholds of 224 Gbps, these tools are indispensable for designing reliable interconnects. The goals of the interconnect analysis are simple – identify interconnects that may fail at the target data rate and allow troubleshooting and fixing the problem. The most efficient approach to such pass-fail analysis is the Decompositional Electromagnetic Analysis (DEA). It allows separation of the signal degradation factors for faster analysis and troubleshooting. This paper outlines the basic elements of DEA, discusses conditions for its accuracy and underscores its significance in the future of interconnect design.**

## I. INTRODUCTION

Data rates in PCB and packaging (PKG) interconnects are increasing in all signaling protocols (PCIe, DDR, Ethernet, USB, SAS, CEI, OIF, UCIe, 5G…). Most of those high-speed signaling standards have one-lane data rates over 6 Gbps (or GT/s) and some up to 224 Gbps with signal spectrum spanning into microwave and mm-wave bandwidths. It is practically impossible to predict behaviour of such interconnects without an accurate electromagnetic (EM) analysis. The goal of such analysis is to identify and fix possible data transmission problems by verifying compliance conditions in frequency and in time domains.

By intent, the interconnects are waveguiding structures and the most efficient way to model them is with the partitioning or decomposition into multiport models of transmission lines and discontinuities. That approach was introduced first for the analysis of closed waveguiding systems [1]. The partitioning was successfully used for design of microwave (MICs) and mm-wave integrated circuits (MMICs) in decompositional EM analysis [2] and in some other design-oriented EM models [3] of open waveguiding systems (microstrip and striplines). The digital PCB/PKG interconnects are also the open waveguiding structures similar to (M)MICs. Though, until early 2000s, the signal integrity (SI) analysis tools used mostly transmission line models (XTK, ICX, HyperLynx). The accuracy of these tools became insufficient as the data rates reached 6 Gbps. Accurate models for such interconnects required analysis that correlates to measurements up to 10-20 GHz. It was not possible to achieve it with the existing SI tools. To address it, the first electromagnetic signal integrity software Simbeor was developed together with a systematic approach to the analysis to measurement validation [4]. The essential elements of the Decompositional Electromagnetic Analysis (DEA) were provided in [5] and the systematic approach was formalized in [6]. The decompositional approach is rigorous field theory technique that allows problem complexity reduction by taking into account the physics of wave propagation. It allows separation of the signal degradation factors that enables faster pass/fail electromagnetic analysis and troubleshooting even on a laptop computer. This paper covers some recent advances in analysis of digital interconnects with DEA.

Partitioning approaches similar to DEA appeared later in the other SI tools (“cut and stitch”, “divide and concur”, “HFSS regions”). However, it is usually misunderstood as an approximate technique. It depends on how it is implemented and what is simulated, as discuss in this paper. EM analysis of interconnects without partitioning was recently enabled by use of formal domain decomposition in finite element methods [7] and formal algebraic matrix reduction in integral equation methods [8], [9] and by availability of high-performance computing. Those are remarkable achievements, but it is also the brute force approach. It is computationally and financially expensive and, thus, not accessible to the majority of the electronics designers who are still relying on design rules or legacy SI software.

## II. DEA TECHNOLOGY EVOLUTION

The DEA is the domain decomposition (DD) technique that is based on the physics of the wave propagation. DEA is the wave approach to the DD. First, structures that can be simulated with transmission line (t-line) models and structures that require 3D EM analysis (discontinuities) are identified. It can be done by following the signal propagation direction between components within some boundaries around signal conductors (waveguiding channel) and setting wave boundaries between all elements as illustrated in Fig. 1. It is a pattern recognition problem that requires fast algorithms to convert polygonal conductors into EM models. All reference conductor discontinuities within the waveguiding channel (pink lines in Fig. 1) are included either in t-line or into discontinuity models. Coupling to the other links within a specified coupling distance can be also included in such model with either coupled t-lines or coupled discontinuities (pads or viaholes). To facilitate the geometry optimization, vertical transitions are identified and parameterized as multi-vias. That allows comparison and model reuse to accelerate the analysis of multiple links (usually, most of the transitions to the same layer are identical). Models for the transmission lines are also compared and re-used. Discontinuity boundaries and boundary conditions, wave ports between the discontinuities and t-lines, external component ports are automatically defined. The transmission line ports are created both for transitions to planar t-lines in the XY-plane and for the Z-directed transitions to BGA and connector pins. The ports in all directions are de-embedded, to eliminate the reflections between the domains. Comparing to the analysis with lumped component ports, the wave ports in all directions extend the accuracy of the DEA to higher frequencies.

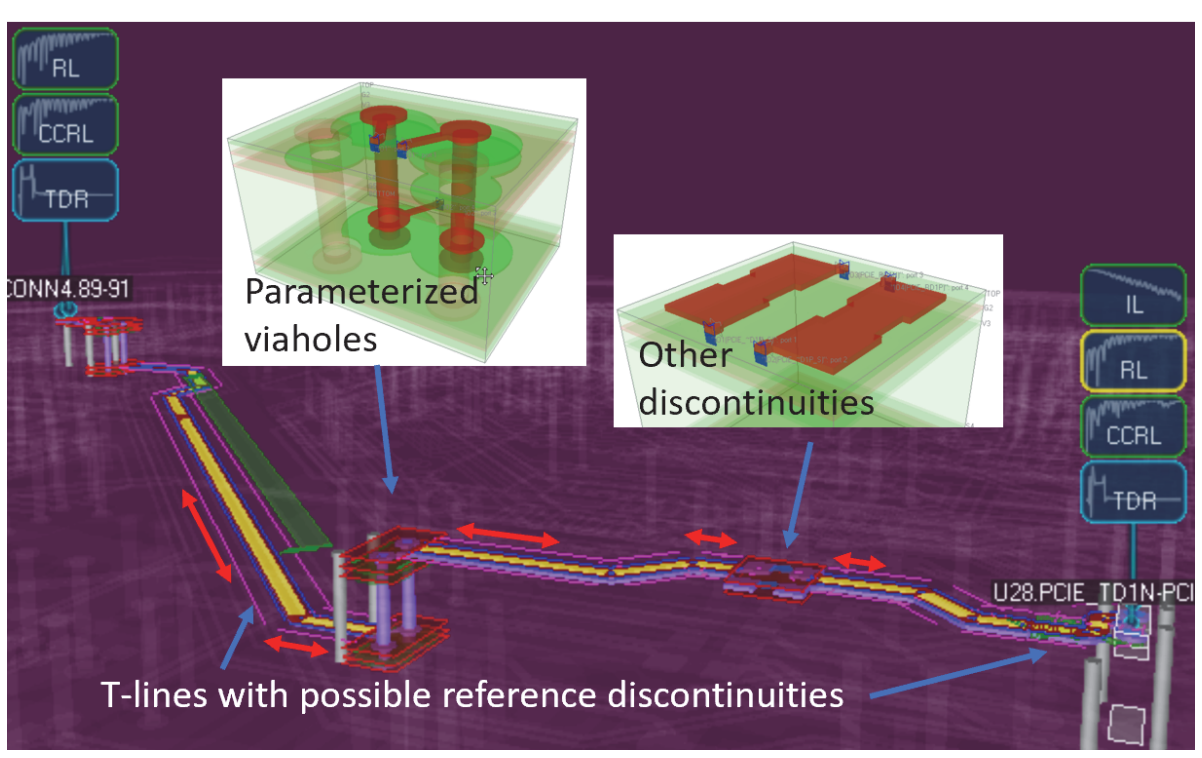

Fig. 1. Differential link decomposition along the waveguiding channel.

The main reason for the decomposition into t-lines and discontinuities is to increase the accuracy and accelerate analysis of long traces. It is computationally expensive to achieve high accuracy in the analysis of long planar t-line segments with the brute force approach due to peculiarities of current distribution in strips and reference planes. It is faster and often more accurate to extract modal characteristic impedances and propagation constants for multiconductor t-lines and use them to model the segments.

The major element of PCB/PKG interconnects are vertical transitions or vias through multi-layered media. Such structures require fast and accurate analysis and optimization. The wave approach to the DD is used to accelerate the analysis of multi-layered discontinuities. The layers are treated as separate domains with the wave channels defined at the partially metallized layers. That turns the matrices describing the whole problem into block band matrices with five block diagonals. Matrices describing blocks are fully populated in the method of lines solver (3DML) and are sparse in the Trefftz finite elements solver (3DTF). The band matrices are solved with the frontal algorithm in both solvers. The complexity of such solution grows linearly with the number of layers in the structure. Overall, that approach accelerates analysis orders of magnitude and substantially reduces the memory requirement.

The analysis of digital interconnects is done in the frequency-domain (FD), but the time-domain (TD) analysis is also required both at the system-level and at the level of each discontinuity. Multiport models require seamless transition between FD and TD. The approximation of S-parameters with rational functions or rational compact models (RCM) enables such analysis in both domains [5]. It is used to accelerate the frequency sweeps by selecting frequencies only where they improve the quality of RCMs. This interpolative sweep reduces the number of frequency points and produces frequency-continuous RCM models for each discontinuity. That allows fast and accurate TD analysis of each discontinuity. At the system-level RCMs are either used directly for both FD and TD analysis or converted into SPICE models, to remove the model dependency on the discretisation and the limited bandwidth.

The DEA analysis acceleration techniques outlined here are implemented in Simbeor software. It enables design exploration and interactive or automated analysis even on a laptop – no HPC is required! To further reduce the simulation time, 3D EM solvers are accelerated with the distributed computing. The analysis of frequency points is parallelized in a local or a cloud network of computers.

## III. INTERCONNECT DESIGN WITH DEA

The interconnects should be designed as the waveguiding structures with the constant characteristic impedance and minimized effect of discontinuities. In reality, it is often not the case. Before running an expensive broadband EM analysis, possible defects should be found and fixed first with inexpensive and fast EM models. This can be done with the multi-pass approach to interconnect design based on a decomposition of signal degradation effects illustrated by the balance of power:

**P_out = P_in - P_abs - P_refl - P_leaked + P_coupled**

Here P_in is the power delivered by a transmitter to the interconnect and P_out is the power delivered to a receiver (degraded signal). All other terms in the balance of power equation describe signal distortion. P_refl characterises the power reflected to the receiver – the major contributors to the reflection are the impedances of traces, pads and viaholes that can be computed and compared with the target impedance at just one frequency point first (signal Nyquist frequency for instance). The problems discovered during such analysis should be addressed in the first pass. P_leaked and P_coupled are two terms that describe a link coupling to neighbour structures (local coupling) as well as to remote structures through parallel-plane waveguides (distant coupling). The structures with possible distant coupling prevent predictability of interconnects with any method and, thus, should be identified and eliminated during the first pass too. It can be done by evaluating localization frequency for each viahole and discontinuity in reference conductors. The structures are predictable with higher confidence over the frequency bandwidth up to the localization frequency. They are also relatively independent from the boundary conditions used to solve the problem in isolation. The leakage from the conditionally localized structures grows with the frequency and the localization frequency can be formally defined as the frequency where P_leaked exceeds some threshold (10% for instance). In cases of high data rates (28-224 Gbps), it may be physically impossible to localized vias on PCBs and we usually need to build models with the bandwidth exceeding the localization limit. Absorption of the lost energy with the low impedance boundary conditions is required to model such structures with the gradual loss of localization. That approximates the behaviour of the power delivery planes and also eliminates the numerical resonances caused by other types of boundary conditions.

During the second pass, the local coupling between parallel traces can be accurately modelled with the coupled transmission line models – such analysis does not require 3D EM models. The local coupling between adjacent vias can be also evaluated separately at this pass. If the local coupling exceeds a coupling compliance condition, it should be fixed.

Only after the impedance, localization and local coupling problems are identified and corrected, more accurate broadband analysis with 3D models of discontinuities is used to evaluate the overall losses caused by material absorption (P_abs) and the reflection (P_refl) over the bandwidth of the signal. Again, if problems are detected, they have to be fixed before the final analysis of multiple links with the local viahole and t-line coupling. If the localization of viaholes is not sufficient, the distant crosstalk can be evaluated and fixed, if necessary. Notice that the key in that multi-pass approach is the localization. The improvement of localization makes interconnects more predictable with any method, including DEA. To support the multi-pass approach to interconnect analysis SI Compliance Analyzer tool was implemented in Simbeor. It allows finding and fixing the problems without resorting to the brute force 3D EM analysis.

## IV. DEA Accuracy and Validation

In context of PCB/PKG interconnect analysis, the accuracy can be defined as a possibility to predict interconnect behaviour. Three major elements define the predictability: localization, broadband material models, and manufacturing adjustments and variations. The localization condition was discussed in the previous section. The broadband dielectrics and conductor roughness models were required even for 10-20 Gbps links [4]. To address this need, the broadband material model identification with GMS-parameters was introduced [10]. It enabled automatic model identification with separation of dielectric and conductor losses and building statistical material models as was demonstrated in [11]. Such models include material and some geometric manufacturing variations and are essential for predictability of 28-224 Gbps interconnects. Possible manufacturing adjustment are also important for the predictability as was demonstrated in [6], where the systematic “sink or swim” approach to the accuracy validation was formally introduced and used to validate Simbeor solvers. As of now, there are no published reports on the systematic validation of the other SI tools. To further formalize and facilitate the validation process, S-parameters similarity metric was introduced in [12].

## V. Conclusion and Future of DEA

Since the formal introduction of DEA for the digital interconnects about 10 years ago [5], it was turned into completely automatic analysis in Simbeor software. That allowed the compliance analysis automation from the design geometry import to the pass-fail report. The outlined multi-pass approach to interconnect design allows elimination of the structures with possible distant coupling and make the final DEA analysis as accurate as necessary. Note that the distant coupling that cannot be avoided can be included in the DEA model by adding additional wave ports between the discontinuities and parallel plane structures.

The outlined DEA opens multiple possibilities to make interconnect compliant by design and to eliminate the need in SI software as it exists now. First, the fast DEA enables the model-driven routing, running analysis simultaneously with the layout. Second, the fast DEA allows analysis of millions of links, to populate solution spaces with all possible geometry and material parameters variations and use of machine learning algorithms to build ranges were interconnects stay compliant as it was demonstrated in [13]. Only the validation of the ranges is required in this approach. Though, for both approaches the process of connecting two components with traces as it is done now in the layout tools should be replace with the process of connecting by waveguiding structures that, in addition to the traces, include relevant reference conductors.